\documentclass[conference,a4paper]{IEEEtran}
\usepackage[utf8]{inputenc} 
\usepackage{units}

\usepackage{tikz,pgfplots}
\usetikzlibrary{calc}
\usetikzlibrary{arrows}
\usetikzlibrary{decorations}
\pgfplotsset{compat=newest}
\pgfplotsset{plot coordinates/math parser=false}
\usetikzlibrary{plotmarks}

\newlength\figureheight
\newlength\figurewidth

\usepackage{cite}

\ifCLASSINFOpdf
\else
\fi
\usepackage{amssymb}
\usepackage[cmex10]{amsmath}
\usepackage{mdwmath}
\usepackage{mdwtab}
\usepackage[caption=false,font=footnotesize]{subfig}
\usepackage{fixltx2e}
\begin{document}
%
\title{Dictionary Adaptation in Sparse Recovery\\Based on Different Types of Coherence}

\author{\IEEEauthorblockN{Henning Zörlein, Faisal Akram and Martin Bossert}
\IEEEauthorblockA{Institute of Communications Engineering\\
Ulm University\\
89081 Ulm, Germany\\
\{henning.zoerlein, martin.bossert\}@uni-ulm.de, faisalakram74@gmail.com}
}


%


\maketitle

\begin{abstract}
In sparse recovery, the unique sparsest solution to an under-determined system of linear equations is of main interest.
This scheme is commonly proposed to be applied to signal acquisition.
In most cases, the signals are not sparse themselves, and therefore, they need to be sparsely represented with the help of a so-called dictionary being specific to the corresponding signal family.
The dictionaries cannot be used for optimization of the resulting under-determined system because they are fixed by the given signal family.
However, the measurement matrix is available for optimization and can be adapted to the dictionary.
Multiple properties of the resulting linear system have been proposed which can be used as objective functions for optimization.
This paper discusses two of them which are both related to the coherence of vectors. 
One property aims for having incoherent measurements, while the other aims for insuring the successful reconstruction.
In the following, the influences of both criteria are compared with different reconstruction approaches.
\end{abstract}


%
\IEEEpeerreviewmaketitle

\section{Introduction}
With the means of sparse recovery, it is possible to obtain the unique sparsest solution to an under-determined system of linear equations, if such a sufficiently sparse solution exists.
This principle is often applied to signal acquisition:
\begin{equation*}
\boldsymbol{y}=\boldsymbol{\Phi}\boldsymbol{x},
\end{equation*}
where the vector $\boldsymbol{x}\in\mathbb{R}^N$ is the unknown signal, the measurement vector $\boldsymbol{y}\in\mathbb{R}^M$ is obtained by applying the measurement matrix $\boldsymbol{\Phi}\in\mathbb{R}^{M\times N}$. 
Although discrete signals are discussed here, the scheme can be extended to the continuous case as well.
Typically, signals of interest are not sparse themselves but can often be sparsely represented with the help of a dictionary $\boldsymbol{\Psi}\in\mathbb{R}^{N\times L}$:
\begin{equation*}
\boldsymbol{x}=\boldsymbol{\Psi}\boldsymbol{\alpha},
\end{equation*}
where $\boldsymbol{\alpha}\in\mathbb{R}^L$ is the sparse representation vector.
With the matrix product $\boldsymbol{\Phi}\boldsymbol{\Psi}=\boldsymbol{A}$, the under-determined system of linear equations
\begin{equation*}
\boldsymbol{y}=\boldsymbol{A}\boldsymbol{\alpha}
\end{equation*}
is obtained for which a sparse solution $\boldsymbol{\hat{\alpha}}$ can be recovered.

The dictionary $\boldsymbol{\Psi}$ is fixed by the signal family under consideration, and therefore, it cannot be used for further optimization.
However, the measurement matrix $\boldsymbol{\Phi}$ is available and can be optimized such that successful sparse recovery is facilitated.
There are different criteria proposed for such optimizations.
In the following, two of them, both based on the coherence of vectors, are compared by their effectiveness.

The remainder of this paper is structured as follows:
In Section~\ref{sec:OptCrit}, the considered criteria and their optimization methods are described.
Afterwards, these schemes are evaluated by numerical simulations in Section~\ref{sec:NumSim}.
Finally, the results of this paper are concluded in Section~\ref{sec:Conclusion}.
\section{Coherence-Based Optimization Criteria}
\label{sec:OptCrit}
It should be noted that the term coherence is differently used in the referred literature. 
In all variants, it refers to the maximal absolute value of the inner product between vectors. 
However, the vectors originate from different sources in the respective publications.

\subsection{Column Coherence of $\boldsymbol{A}$}
The mutual column coherence $\mu(\boldsymbol{A})$ describes the maximal absolute value of the inner product between all columns of $\boldsymbol{A}$:

\begin{equation*}
\mu(\boldsymbol{A}) = \max \limits_{i\neq j} \frac{\mid\langle\boldsymbol{a}_i,\boldsymbol{a}_j\rangle\mid}{{{\|\boldsymbol{a}_i\|}_2}{{\|\boldsymbol{a}_j\|}_2}},
\end{equation*}
where $\boldsymbol{a}_i$ is the $i$-th column of $\boldsymbol{A}$.
Multiple conditions on $\mu(\boldsymbol{A})$ have been proposed guaranteeing the successful recovery for a certain sparsity or the existence of a unique sparsest solution ~\cite{DonohoHuoNov01,DonohoEladDec02,EladBrucksteinSep02,GribonvalNielsenDec03}.
Additionally, reconstruction algorithms like Orthogonal Matching Pursuit (OMP)~\cite{Pati93orthogonalmatching} rely on a low mutual column coherence, and therefore, gain especially from its optimization.
Thus, the usage of this property is highly motivated.

Such conditions on the coherence are often used for scenarios, where the signal vector $\boldsymbol{x}$ is sparse itself (which corresponds to the case of a dictionary being the identity matrix $\boldsymbol{\Psi}=\boldsymbol{\mathrm{I}}$). 
However, this property is also important for arbitrary dictionaries.
There are several approaches to optimize $\boldsymbol{\Phi}$ such that the columns of $\boldsymbol{A}$ are less coherent, e.g.~\cite{Elad07,DuarteCarvajalinoSapiro2009,AbolghasemiFerdowsiMakkiabadiSanei2010,AbolghasemiJarchiSanei2010,XuPiCao2010,TsiligianniKondiKatsaggelos2012}.
In the remainder of this paper, the approach by Elad given in~\cite{Elad07} is considered for optimizing $\mu(\boldsymbol{A})$.

\subsection{Row Coherence of $\boldsymbol{\Phi}$ with respect to Columns of $\boldsymbol{\Psi}$}
Especially for the scenario of non-sparse signals, another kind of coherence between $\boldsymbol{\Phi}$ and $\boldsymbol{\Psi}$ is discussed in~\cite{CandesRomberg2007}:
\begin{equation*}
\mu(\boldsymbol{\Phi},\boldsymbol{\Psi}) = \max \limits_{i,j} \frac{\mid\langle\boldsymbol{\phi}_i,\boldsymbol{\psi}_j\rangle\mid}{{{\|\boldsymbol{\phi}_i\|}_2}{{\|\boldsymbol{\psi}_j\|}_2}},
\end{equation*}
where $\boldsymbol{\phi}_i$ is the $i$-th row of $\boldsymbol{\Phi}$ and $\boldsymbol{\psi}_j$ is the $j$-th column of~$\boldsymbol{\Psi}$.

This coherence property is motivated by the measurement process: A small value of $\mu(\boldsymbol{\Phi},\boldsymbol{\Psi})$ results in measurements being more independent and equally adapted to all $\boldsymbol{\psi}_j$.

The optimization method given in~\cite{LazZoerBoss:SCC2013} is slightly adapted to be used here in order to obtain measurement matrices $\boldsymbol{\Phi}$ optimized with respect to $\mu(\boldsymbol{\Phi},\boldsymbol{\Psi})$.
\section{Numerical Evaluation}
\label{sec:NumSim}
In our simulations, measurement matrices $\boldsymbol{\Phi}_{\mu(\boldsymbol{A})}$ and $\boldsymbol{\Phi}_{\mu(\boldsymbol{\Phi},\boldsymbol{\Psi})}$ (optimized according to $\mu(\boldsymbol{A})$ and $\mu(\boldsymbol{\Phi},\boldsymbol{\Psi})$ respectively) are compared.
As reference, a column-normalized Gaussian random measurement matrices $\boldsymbol{\Phi}_{\textrm{rand}}$ is considered as well.
We used a concatenation of identity and Discrete Cosine Transform (DCT) matrix $\boldsymbol{\Psi}_{\left[\boldsymbol{\mathrm{I}},\boldsymbol{\mathrm{DCT}}\right]}$, and a column-normalized Gaussian random matrix  $\boldsymbol{\Psi}_{\textrm{rand}}$, as dictionaries.

\subsection{Verify Success of Optimizations}
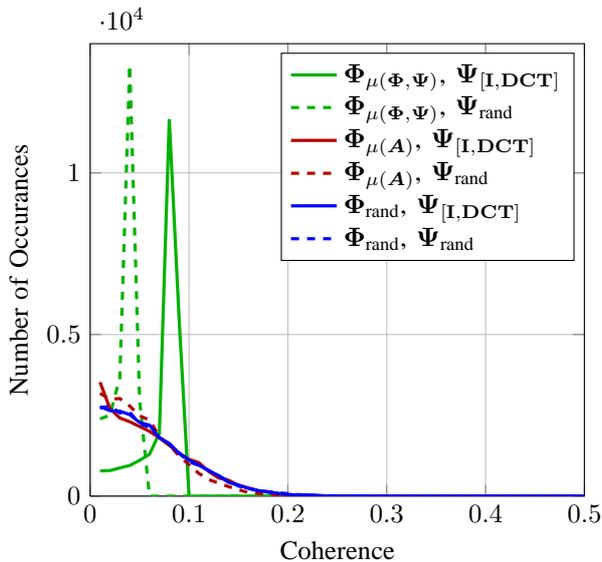
\begin{figure}
\centering
\setlength\figureheight{6cm}
\setlength\figurewidth{6.5cm}
%
%
%
%
\begin{tikzpicture}

\begin{axis}[%
width=\figurewidth,
height=\figureheight,
scale only axis,
xlabel={Coherence},
ylabel={Number of Occurances},
xmin=0, xmax=0.5,
ymin=0, ymax=14000,
ymajorgrids,
xmajorgrids,
legend style={draw=black,fill=white,legend cell align=left}]
\addplot [
color=green!70!black,
very thick,
solid
]
table{
0.01 784
0.02 790
0.03 874
0.04 948
0.05 1108
0.06 1296
0.07 1950
0.08 11664
0.09 5456
0.1 0
0.11 0
0.12 0
0.13 0
0.14 0
0.15 0
0.16 0
0.17 0
0.18 0
0.19 0
0.2 0
0.21 0
0.22 0
0.23 0
0.24 0
0.25 0
0.26 0
0.27 0
0.28 0
0.29 0
0.3 0
0.31 0
0.32 0
0.33 0
0.34 0
0.35 0
0.36 0
0.37 0
0.38 0
0.39 0
0.4 0
0.41 0
0.42 0
0.43 0
0.44 0
0.45 0
0.46 0
0.47 0
0.48 0
0.49 0
0.5 0
0.51 0
0.52 0
0.53 0
0.54 0
0.55 0
0.56 0
0.57 0
0.58 0
0.59 0
0.6 0
0.61 0
0.62 0
0.63 0
0.64 0
0.65 0
0.66 0
0.67 0
0.68 0
0.69 0
0.7 0
0.71 0
0.72 0
0.73 0
0.74 0
0.75 0
0.76 0
0.77 0
0.78 0
0.79 0
0.8 0
0.81 0
0.82 0
0.83 0
0.84 0
0.85 0
0.86 0
0.87 0
0.88 0
0.89 0
0.9 0
0.91 0
0.92 0
0.93 0
0.94 0
0.95 0
0.96 0
0.97 0
0.98 0
0.99 0
1 0
};
\addlegendentry{$\boldsymbol{\Phi}_{\mu(\boldsymbol{\Phi},\boldsymbol{\Psi})}$, $\boldsymbol{\Psi}_{\left[\boldsymbol{\mathrm{I}},\boldsymbol{\mathrm{DCT}}\right]}$};

\addplot [
color=green!70!black,
very thick,
dashed
]
table{
0.01 2390
0.02 2504
0.03 3672
0.04 13340
0.05 2964
0.06 0
0.07 0
0.08 0
0.09 0
0.1 0
0.11 0
0.12 0
0.13 0
0.14 0
0.15 0
0.16 0
0.17 0
0.18 0
0.19 0
0.2 0
0.21 0
0.22 0
0.23 0
0.24 0
0.25 0
0.26 0
0.27 0
0.28 0
0.29 0
0.3 0
0.31 0
0.32 0
0.33 0
0.34 0
0.35 0
0.36 0
0.37 0
0.38 0
0.39 0
0.4 0
0.41 0
0.42 0
0.43 0
0.44 0
0.45 0
0.46 0
0.47 0
0.48 0
0.49 0
0.5 0
0.51 0
0.52 0
0.53 0
0.54 0
0.55 0
0.56 0
0.57 0
0.58 0
0.59 0
0.6 0
0.61 0
0.62 0
0.63 0
0.64 0
0.65 0
0.66 0
0.67 0
0.68 0
0.69 0
0.7 0
0.71 0
0.72 0
0.73 0
0.74 0
0.75 0
0.76 0
0.77 0
0.78 0
0.79 0
0.8 0
0.81 0
0.82 0
0.83 0
0.84 0
0.85 0
0.86 0
0.87 0
0.88 0
0.89 0
0.9 0
0.91 0
0.92 0
0.93 0
0.94 0
0.95 0
0.96 0
0.97 0
0.98 0
0.99 0
1 0
};
\addlegendentry{$\boldsymbol{\Phi}_{\mu(\boldsymbol{\Phi},\boldsymbol{\Psi})}$, $\boldsymbol{\Psi}_{\textrm{rand}}$};

\addplot [
color=red!70!black,
very thick,
solid
]
table{
0.01 3524
0.02 2712
0.03 2418
0.04 2302
0.05 2154
0.06 2006
0.07 1802
0.08 1570
0.09 1328
0.1 1138
0.11 1034
0.12 762
0.13 562
0.14 492
0.15 334
0.16 254
0.17 160
0.18 124
0.19 70
0.2 44
0.21 34
0.22 16
0.23 12
0.24 8
0.25 6
0.26 4
0.27 0
0.28 0
0.29 0
0.3 0
0.31 0
0.32 0
0.33 0
0.34 0
0.35 0
0.36 0
0.37 0
0.38 0
0.39 0
0.4 0
0.41 0
0.42 0
0.43 0
0.44 0
0.45 0
0.46 0
0.47 0
0.48 0
0.49 0
0.5 0
0.51 0
0.52 0
0.53 0
0.54 0
0.55 0
0.56 0
0.57 0
0.58 0
0.59 0
0.6 0
0.61 0
0.62 0
0.63 0
0.64 0
0.65 0
0.66 0
0.67 0
0.68 0
0.69 0
0.7 0
0.71 0
0.72 0
0.73 0
0.74 0
0.75 0
0.76 0
0.77 0
0.78 0
0.79 0
0.8 0
0.81 0
0.82 0
0.83 0
0.84 0
0.85 0
0.86 0
0.87 0
0.88 0
0.89 0
0.9 0
0.91 0
0.92 0
0.93 0
0.94 0
0.95 0
0.96 0
0.97 0
0.98 0
0.99 0
1 0
};
\addlegendentry{$\boldsymbol{\Phi}_{\mu(\boldsymbol{A})}$, $\boldsymbol{\Psi}_{\left[\boldsymbol{\mathrm{I}},\boldsymbol{\mathrm{DCT}}\right]}$};

\addplot [
color=red!70!black,
very thick,
dashed
]
table{
0.01 3178
0.02 3000
0.03 3020
0.04 2788
0.05 2484
0.06 2352
0.07 1844
0.08 1582
0.09 1246
0.1 974
0.11 702
0.12 512
0.13 402
0.14 294
0.15 200
0.16 112
0.17 84
0.18 54
0.19 26
0.2 8
0.21 6
0.22 2
0.23 0
0.24 0
0.25 0
0.26 0
0.27 0
0.28 0
0.29 0
0.3 0
0.31 0
0.32 0
0.33 0
0.34 0
0.35 0
0.36 0
0.37 0
0.38 0
0.39 0
0.4 0
0.41 0
0.42 0
0.43 0
0.44 0
0.45 0
0.46 0
0.47 0
0.48 0
0.49 0
0.5 0
0.51 0
0.52 0
0.53 0
0.54 0
0.55 0
0.56 0
0.57 0
0.58 0
0.59 0
0.6 0
0.61 0
0.62 0
0.63 0
0.64 0
0.65 0
0.66 0
0.67 0
0.68 0
0.69 0
0.7 0
0.71 0
0.72 0
0.73 0
0.74 0
0.75 0
0.76 0
0.77 0
0.78 0
0.79 0
0.8 0
0.81 0
0.82 0
0.83 0
0.84 0
0.85 0
0.86 0
0.87 0
0.88 0
0.89 0
0.9 0
0.91 0
0.92 0
0.93 0
0.94 0
0.95 0
0.96 0
0.97 0
0.98 0
0.99 0
1 0
};
\addlegendentry{$\boldsymbol{\Phi}_{\mu(\boldsymbol{A})}$, $\boldsymbol{\Psi}_{\textrm{rand}}$};

\addplot [
color=blue,
very thick,
solid
]
table{
0.01 2768
0.02 2646
0.03 2628
0.04 2522
0.05 2280
0.06 2196
0.07 1804
0.08 1632
0.09 1326
0.1 1116
0.11 956
0.12 784
0.13 638
0.14 424
0.15 342
0.16 240
0.17 174
0.18 134
0.19 88
0.2 52
0.21 34
0.22 40
0.23 22
0.24 4
0.25 4
0.26 6
0.27 6
0.28 2
0.29 2
0.3 0
0.31 0
0.32 0
0.33 0
0.34 0
0.35 0
0.36 0
0.37 0
0.38 0
0.39 0
0.4 0
0.41 0
0.42 0
0.43 0
0.44 0
0.45 0
0.46 0
0.47 0
0.48 0
0.49 0
0.5 0
0.51 0
0.52 0
0.53 0
0.54 0
0.55 0
0.56 0
0.57 0
0.58 0
0.59 0
0.6 0
0.61 0
0.62 0
0.63 0
0.64 0
0.65 0
0.66 0
0.67 0
0.68 0
0.69 0
0.7 0
0.71 0
0.72 0
0.73 0
0.74 0
0.75 0
0.76 0
0.77 0
0.78 0
0.79 0
0.8 0
0.81 0
0.82 0
0.83 0
0.84 0
0.85 0
0.86 0
0.87 0
0.88 0
0.89 0
0.9 0
0.91 0
0.92 0
0.93 0
0.94 0
0.95 0
0.96 0
0.97 0
0.98 0
0.99 0
1 0
};
\addlegendentry{$\boldsymbol{\Phi}_{\textrm{rand}}$, $\boldsymbol{\Psi}_{\left[\boldsymbol{\mathrm{I}},\boldsymbol{\mathrm{DCT}}\right]}$};

\addplot [
color=blue,
very thick,
dashed
]
table{
0.01 2726
0.02 2784
0.03 2578
0.04 2592
0.05 2244
0.06 2132
0.07 1824
0.08 1614
0.09 1328
0.1 1062
0.11 952
0.12 792
0.13 618
0.14 470
0.15 312
0.16 240
0.17 170
0.18 130
0.19 104
0.2 76
0.21 48
0.22 34
0.23 8
0.24 12
0.25 8
0.26 8
0.27 2
0.28 2
0.29 0
0.3 0
0.31 0
0.32 0
0.33 0
0.34 0
0.35 0
0.36 0
0.37 0
0.38 0
0.39 0
0.4 0
0.41 0
0.42 0
0.43 0
0.44 0
0.45 0
0.46 0
0.47 0
0.48 0
0.49 0
0.5 0
0.51 0
0.52 0
0.53 0
0.54 0
0.55 0
0.56 0
0.57 0
0.58 0
0.59 0
0.6 0
0.61 0
0.62 0
0.63 0
0.64 0
0.65 0
0.66 0
0.67 0
0.68 0
0.69 0
0.7 0
0.71 0
0.72 0
0.73 0
0.74 0
0.75 0
0.76 0
0.77 0
0.78 0
0.79 0
0.8 0
0.81 0
0.82 0
0.83 0
0.84 0
0.85 0
0.86 0
0.87 0
0.88 0
0.89 0
0.9 0
0.91 0
0.92 0
0.93 0
0.94 0
0.95 0
0.96 0
0.97 0
0.98 0
0.99 0
1 0
};
\addlegendentry{$\boldsymbol{\Phi}_{\textrm{rand}}$, $\boldsymbol{\Psi}_{\textrm{rand}}$};

\end{axis}
\end{tikzpicture}%
\caption{\label{fig:PhiTPsi_EyeDCT}Coherence distribution of $\left[\boldsymbol{\Phi}^T,\boldsymbol{\Psi}\right]$ for $M=30$, $N=200$ and $L=400$. Intra column coherence of $\boldsymbol{\Psi}$ is removed.}
\end{figure}

The coherence distribution of $\left[\boldsymbol{\Phi}^T,\boldsymbol{\Psi}\right]$ can be used to evaluate the success of a $\mu(\boldsymbol{\Phi},\boldsymbol{\Psi})$ optimization.
In Fig.~\ref{fig:PhiTPsi_EyeDCT}, the intra column coherence of $\boldsymbol{\Psi}$, which cannot be changed, is removed in order to evaluate the optimization results.
As expected, the coherence distributions of $\boldsymbol{\Phi}_{\mu(\boldsymbol{\Phi},\boldsymbol{\Psi})}$ show the effect of the optimization, whilst $\boldsymbol{\Phi}_{\mu(\boldsymbol{A})}$ performs as $\boldsymbol{\Phi}_{\textrm{rand}}$.
Therefore, the successful optimization is verified.

\begin{figure}
\centering
\setlength\figureheight{6cm}
\setlength\figurewidth{6.5cm}
%
%
%
%
\begin{tikzpicture}

\begin{axis}[%
width=\figurewidth,
height=\figureheight,
scale only axis,
xlabel={Coherence},
ylabel={Number of Occurances},
xmin=0, xmax=1,
ymin=0, ymax=10000,
ymajorgrids,
xmajorgrids,
legend style={draw=black,fill=white,legend cell align=left}]
\addplot [
color=green!70!black,
very thick,
solid
]
table{
0.01 7258
0.02 7008
0.03 6974
0.04 6896
0.05 6796
0.06 6762
0.07 6682
0.08 6318
0.09 6194
0.1 6022
0.11 5944
0.12 5514
0.13 5486
0.14 5468
0.15 5110
0.16 4948
0.17 4572
0.18 4572
0.19 4062
0.2 3956
0.21 3728
0.22 3540
0.23 3218
0.24 3006
0.25 2832
0.26 2580
0.27 2412
0.28 2278
0.29 2128
0.3 1876
0.31 1726
0.32 1554
0.33 1378
0.34 1364
0.35 1104
0.36 994
0.37 886
0.38 882
0.39 758
0.4 698
0.41 580
0.42 474
0.43 364
0.44 410
0.45 288
0.46 344
0.47 248
0.48 218
0.49 188
0.5 178
0.51 138
0.52 124
0.53 80
0.54 78
0.55 74
0.56 54
0.57 38
0.58 36
0.59 50
0.6 10
0.61 30
0.62 16
0.63 14
0.64 18
0.65 12
0.66 8
0.67 14
0.68 6
0.69 2
0.7 8
0.71 4
0.72 0
0.73 4
0.74 4
0.75 0
0.76 0
0.77 0
0.78 0
0.79 0
0.8 0
0.81 0
0.82 0
0.83 0
0.84 2
0.85 0
0.86 0
0.87 0
0.88 0
0.89 0
0.9 0
0.91 0
0.92 0
0.93 0
0.94 0
0.95 0
0.96 0
0.97 0
0.98 0
0.99 0
1 0
};
\addlegendentry{$\boldsymbol{\Phi}_{\mu(\boldsymbol{\Phi},\boldsymbol{\Psi})}$, $\boldsymbol{\Psi}_{\left[\boldsymbol{\mathrm{I}},\boldsymbol{\mathrm{DCT}}\right]}$};

\addplot [
color=green!70!black,
very thick,
dashed
]
table{
0.01 7120
0.02 6884
0.03 7100
0.04 6846
0.05 6962
0.06 6760
0.07 6790
0.08 6360
0.09 6098
0.1 5996
0.11 6200
0.12 5772
0.13 5784
0.14 5318
0.15 5182
0.16 4720
0.17 4646
0.18 4424
0.19 4286
0.2 3846
0.21 3802
0.22 3398
0.23 3298
0.24 3154
0.25 2830
0.26 2616
0.27 2336
0.28 2298
0.29 2040
0.3 1978
0.31 1644
0.32 1588
0.33 1378
0.34 1270
0.35 1052
0.36 984
0.37 900
0.38 854
0.39 700
0.4 642
0.41 538
0.42 472
0.43 408
0.44 344
0.45 328
0.46 266
0.47 206
0.48 192
0.49 166
0.5 146
0.51 106
0.52 104
0.53 76
0.54 70
0.55 50
0.56 66
0.57 26
0.58 26
0.59 24
0.6 34
0.61 34
0.62 14
0.63 14
0.64 12
0.65 6
0.66 8
0.67 0
0.68 4
0.69 0
0.7 2
0.71 0
0.72 2
0.73 0
0.74 0
0.75 0
0.76 0
0.77 0
0.78 0
0.79 0
0.8 0
0.81 0
0.82 0
0.83 0
0.84 0
0.85 0
0.86 0
0.87 0
0.88 0
0.89 0
0.9 0
0.91 0
0.92 0
0.93 0
0.94 0
0.95 0
0.96 0
0.97 0
0.98 0
0.99 0
1 0
};
\addlegendentry{$\boldsymbol{\Phi}_{\mu(\boldsymbol{\Phi},\boldsymbol{\Psi})}$, $\boldsymbol{\Psi}_{\textrm{rand}}$};

\addplot [
color=red!70!black,
very thick,
solid
]
table{
0.01 3802
0.02 3838
0.03 3990
0.04 3964
0.05 4062
0.06 4006
0.07 4080
0.08 4250
0.09 4422
0.1 4582
0.11 5048
0.12 7834
0.13 8856
0.14 9174
0.15 9216
0.16 8920
0.17 8598
0.18 8024
0.19 7210
0.2 6312
0.21 4608
0.22 4064
0.23 3476
0.24 3384
0.25 2852
0.26 2642
0.27 2350
0.28 2096
0.29 1900
0.3 1632
0.31 1424
0.32 1210
0.33 1072
0.34 1016
0.35 838
0.36 726
0.37 658
0.38 544
0.39 454
0.4 366
0.41 390
0.42 278
0.43 224
0.44 212
0.45 192
0.46 148
0.47 94
0.48 122
0.49 64
0.5 76
0.51 64
0.52 46
0.53 44
0.54 24
0.55 18
0.56 28
0.57 16
0.58 16
0.59 6
0.6 2
0.61 6
0.62 2
0.63 4
0.64 4
0.65 2
0.66 2
0.67 8
0.68 2
0.69 0
0.7 0
0.71 6
0.72 0
0.73 0
0.74 0
0.75 0
0.76 0
0.77 0
0.78 0
0.79 0
0.8 0
0.81 0
0.82 0
0.83 0
0.84 0
0.85 0
0.86 0
0.87 0
0.88 0
0.89 0
0.9 0
0.91 0
0.92 0
0.93 0
0.94 0
0.95 0
0.96 0
0.97 0
0.98 0
0.99 0
1 0
};
\addlegendentry{$\boldsymbol{\Phi}_{\mu(\boldsymbol{A})}$, $\boldsymbol{\Psi}_{\left[\boldsymbol{\mathrm{I}},\boldsymbol{\mathrm{DCT}}\right]}$};

\addplot [
color=red!70!black,
very thick,
dashed
]
table{
0.01 3900
0.02 3946
0.03 3894
0.04 4066
0.05 3816
0.06 3876
0.07 4258
0.08 4190
0.09 4294
0.1 4630
0.11 5134
0.12 7796
0.13 8712
0.14 9074
0.15 9202
0.16 9022
0.17 8552
0.18 7980
0.19 7314
0.2 6362
0.21 4718
0.22 3856
0.23 3538
0.24 3288
0.25 3018
0.26 2650
0.27 2352
0.28 2026
0.29 1798
0.3 1588
0.31 1388
0.32 1332
0.33 1114
0.34 1058
0.35 844
0.36 758
0.37 696
0.38 536
0.39 456
0.4 394
0.41 372
0.42 300
0.43 242
0.44 216
0.45 150
0.46 164
0.47 128
0.48 92
0.49 92
0.5 66
0.51 62
0.52 50
0.53 54
0.54 32
0.55 26
0.56 16
0.57 24
0.58 14
0.59 8
0.6 8
0.61 12
0.62 6
0.63 8
0.64 4
0.65 6
0.66 4
0.67 2
0.68 2
0.69 2
0.7 4
0.71 0
0.72 2
0.73 2
0.74 0
0.75 2
0.76 0
0.77 0
0.78 0
0.79 2
0.8 0
0.81 0
0.82 0
0.83 0
0.84 0
0.85 0
0.86 0
0.87 0
0.88 0
0.89 0
0.9 0
0.91 0
0.92 0
0.93 0
0.94 0
0.95 0
0.96 0
0.97 0
0.98 0
0.99 0
1 0
};
\addlegendentry{$\boldsymbol{\Phi}_{\mu(\boldsymbol{A})}$, $\boldsymbol{\Psi}_{\textrm{rand}}$};

\addplot [
color=blue,
very thick,
solid
]
table{
0.01 6688
0.02 6370
0.03 6350
0.04 6382
0.05 6492
0.06 6432
0.07 6342
0.08 6246
0.09 5834
0.1 5856
0.11 5574
0.12 5460
0.13 5470
0.14 5174
0.15 5158
0.16 4934
0.17 4660
0.18 4464
0.19 4178
0.2 4142
0.21 3862
0.22 3692
0.23 3348
0.24 3186
0.25 3070
0.26 2954
0.27 2674
0.28 2508
0.29 2174
0.3 2076
0.31 1966
0.32 1768
0.33 1564
0.34 1430
0.35 1342
0.36 1272
0.37 1090
0.38 970
0.39 892
0.4 768
0.41 702
0.42 582
0.43 556
0.44 408
0.45 416
0.46 344
0.47 304
0.48 212
0.49 206
0.5 196
0.51 154
0.52 170
0.53 102
0.54 86
0.55 64
0.56 44
0.57 58
0.58 52
0.59 30
0.6 16
0.61 14
0.62 20
0.63 16
0.64 12
0.65 6
0.66 2
0.67 6
0.68 6
0.69 0
0.7 4
0.71 0
0.72 0
0.73 0
0.74 0
0.75 0
0.76 0
0.77 0
0.78 0
0.79 0
0.8 0
0.81 0
0.82 0
0.83 0
0.84 0
0.85 0
0.86 0
0.87 0
0.88 0
0.89 0
0.9 0
0.91 0
0.92 0
0.93 0
0.94 0
0.95 0
0.96 0
0.97 0
0.98 0
0.99 0
1 0
};
\addlegendentry{$\boldsymbol{\Phi}_{\textrm{rand}}$, $\boldsymbol{\Psi}_{\left[\boldsymbol{\mathrm{I}},\boldsymbol{\mathrm{DCT}}\right]}$};

\addplot [
color=blue,
very thick,
dashed
]
table{
0.01 6466
0.02 6460
0.03 6194
0.04 6468
0.05 6338
0.06 6036
0.07 5966
0.08 6104
0.09 6022
0.1 5896
0.11 5584
0.12 5378
0.13 5134
0.14 5144
0.15 4994
0.16 4790
0.17 4624
0.18 4310
0.19 4080
0.2 4084
0.21 4004
0.22 3568
0.23 3526
0.24 3298
0.25 3090
0.26 2824
0.27 2744
0.28 2494
0.29 2408
0.3 2120
0.31 1922
0.32 1796
0.33 1752
0.34 1564
0.35 1472
0.36 1342
0.37 1182
0.38 974
0.39 970
0.4 860
0.41 854
0.42 628
0.43 574
0.44 496
0.45 498
0.46 438
0.47 346
0.48 284
0.49 266
0.5 266
0.51 192
0.52 144
0.53 114
0.54 94
0.55 74
0.56 80
0.57 50
0.58 50
0.59 52
0.6 28
0.61 20
0.62 18
0.63 24
0.64 6
0.65 4
0.66 2
0.67 4
0.68 4
0.69 4
0.7 0
0.71 0
0.72 2
0.73 0
0.74 0
0.75 0
0.76 2
0.77 0
0.78 0
0.79 0
0.8 0
0.81 0
0.82 0
0.83 0
0.84 0
0.85 0
0.86 0
0.87 0
0.88 0
0.89 0
0.9 0
0.91 0
0.92 0
0.93 0
0.94 0
0.95 0
0.96 0
0.97 0
0.98 0
0.99 0
1 0
};
\addlegendentry{$\boldsymbol{\Phi}_{\textrm{rand}}$, $\boldsymbol{\Psi}_{\textrm{rand}}$};

\end{axis}
\end{tikzpicture}%
\caption{\label{fig:CoherOfA}Coherence distribution of $\boldsymbol{A}$ for $M=30$, $N=200$ and $L=400$.}
\end{figure}
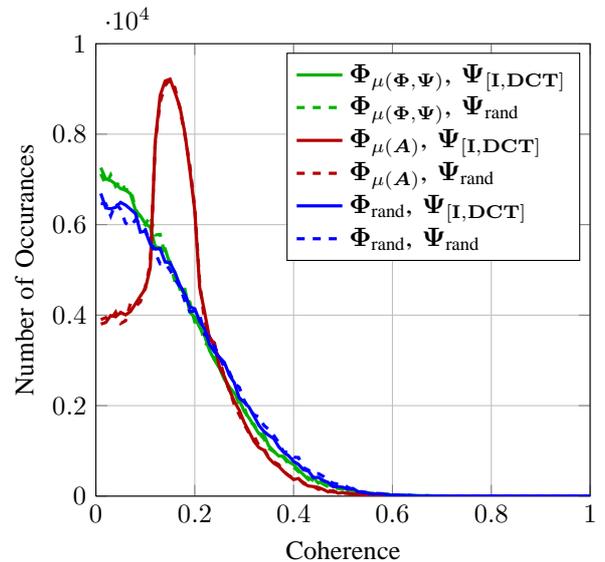

The success of a optimization with respect to $\mu(\boldsymbol{A})$, can be observed by the coherence distribution of $\boldsymbol{A}$ shown in Fig.~\ref{fig:CoherOfA}.
It is successful and independent of the used dictionary, c.f.~\cite{Elad07}.
The measurement matrix $\boldsymbol{\Phi}_{\mu(\boldsymbol{\Phi},\boldsymbol{\Psi})}$ results in a coherence distribution similar to the same for $\boldsymbol{\Phi}_{\textrm{rand}}$.


\subsection{Evaluation of Effectiveness}
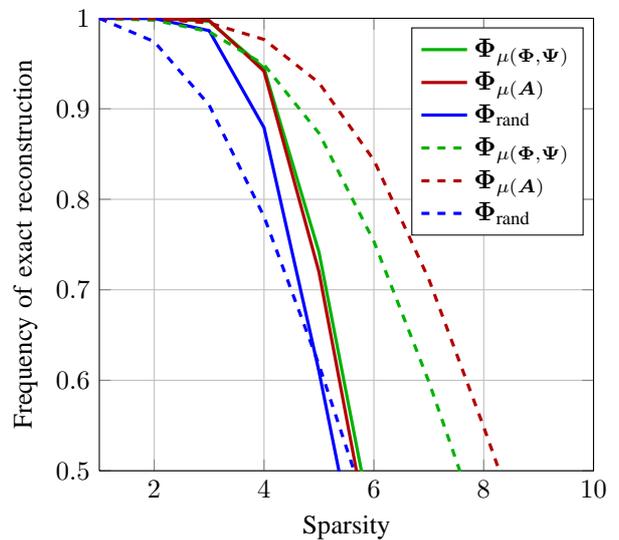
\begin{figure}
\centering
\setlength\figureheight{6cm}
\setlength\figurewidth{6.5cm}
%
%
%
%
\begin{tikzpicture}

\begin{axis}[%
width=\figurewidth,
height=\figureheight,
scale only axis,
xlabel={Sparsity},
ylabel={Frequency of exact reconstruction},
xmin=1, xmax=10,
ymin=0.5, ymax=1,
ymajorgrids,
xmajorgrids,
legend style={draw=black,fill=white,legend cell align=left}]
\addplot [
color=green!70!black,
very thick,
solid
]
table{
1 1
2 0.99999
3 0.99671
4 0.94497
5 0.74165
6 0.4289
};
\addlegendentry{$\boldsymbol{\Phi}_{\mu(\boldsymbol{\Phi},\boldsymbol{\Psi})}$};

\addplot [
color=red!70!black,
very thick,
solid
]
table{
1 1
2 0.99998
3 0.99775
4 0.94155
5 0.71967
6 0.39773
};
\addlegendentry{$\boldsymbol{\Phi}_{\mu(\boldsymbol{A})}$};

\addplot [
color=blue,
very thick,
solid
]
table{
1 1
2 0.99979
3 0.98635
4 0.87919
5 0.60986
6 0.3074
};
\addlegendentry{$\boldsymbol{\Phi}_{\text{rand}}$};

\addplot [
color=green!70!black,
very thick,
dashed
]
table{
1 1
2 0.99789
3 0.98527
4 0.94881
5 0.87302
6 0.75343
7 0.59864
8 0.42398
};
\addlegendentry{$\boldsymbol{\Phi}_{\mu(\boldsymbol{\Phi},\boldsymbol{\Psi})}$};

\addplot [
color=red!70!black,
very thick,
dashed
]
table{
1 1
2 0.99955
3 0.9949
4 0.97661
5 0.92873
6 0.84319
7 0.7118
8 0.54844
9 0.37763
};
\addlegendentry{$\boldsymbol{\Phi}_{\mu(\boldsymbol{A})}$};

\addplot [
color=blue,
very thick,
dashed
]
table{
1 1
2 0.97423
3 0.90414
4 0.78087
5 0.61713
6 0.43376
};
\addlegendentry{$\boldsymbol{\Phi}_{\text{rand}}$};

\end{axis}
\end{tikzpicture}%
\caption{\label{fig:fexact}Frequency of exact reconstruction with $\boldsymbol{\Psi}_{\left[\boldsymbol{\mathrm{I}},\boldsymbol{\mathrm{DCT}}\right]}$ for $M=30$, $N=200$ and $L=400$. BP is used for solid lines, OMP for dashed lines.}
\end{figure}

The frequency of exact reconstruction over the sparsity is used to determine which optimization approach is more effective.
A reconstruction is considered to be successful, if
$
\left\| \boldsymbol{\Psi \alpha}-\boldsymbol{\Psi \hat{\alpha}}\right\|_2<10^{-3}.
$
We performed $100000$ iterations for each sparsity level, where we used Basis Pursuit (BP)~\cite{ChenDonohoSaunders98} and OMP as reconstruction algorithms.
The result for the dictionary $\boldsymbol{\Psi}_{\left[\boldsymbol{\mathrm{I}},\boldsymbol{\mathrm{DCT}}\right]}$ is shown in Fig.~\ref{fig:fexact}.
Both optimization approaches lead to an increased efficiency, where OMP gains especially for $\boldsymbol{\Phi}_{\mu(\boldsymbol{A})}$, as it can be seen in the figure.
For BP, the approach of $\boldsymbol{\Phi}_{\mu(\boldsymbol{\Phi},\boldsymbol{\Psi})}$ is slightly better for almost all sparsity levels.
Thus, in case of reconstruction algorithms relying on the column coherence of the sensing matrix $\boldsymbol{A}$, approaches aiming for low values of $\mu(\boldsymbol{A})$ are naturally superior.
However, for other reconstruction algorithms, $\mu(\boldsymbol{\Phi},\boldsymbol{\Psi})$ should be considered as optimization objective as well.
For other dictionaries (e.g.~$\boldsymbol{\Psi}_{\textrm{rand}}$) similar results have been obtained.

\section{Conclusion}
\label{sec:Conclusion}
In our simulations, we showed that it is more effective to optimize $\mu(\boldsymbol{A})$ rather than $\mu(\boldsymbol{\Phi},\boldsymbol{\Psi})$ for the adaptation of the sensing matrix $\boldsymbol{\Phi}$ with respect to a given dictionary~$\boldsymbol{\Psi}$ in case of reconstruction algorithms which rely on the column coherence of $\boldsymbol{A}$, like OMP.
For other reconstructions algorithms, the optimization of $\mu(\boldsymbol{\Phi},\boldsymbol{\Psi})$ might also be considered.


\section*{Acknowledgments}

This work was supported by the German research council Deutsche Forschungsgemeinschaft (DFG) under Grant Bo~867/27-1.



%

\bibliographystyle{IEEEtran}
\bibliography{CoSeRa2013}

\end{document}